# Thermoelectric La-doped $SrTiO_3$ epitaxial layers with single-crystal quality: from nm to μm and mosaicity effects

M. Apreutesei[a], R. Debord[b], M. Bouras[a], P. Regreny[a], C. Botella[a], A. Benamrouche[a], A. Carretero-Genevrier[a,c], J. Gazquez[d], G. Grenet[a], S. Pailhès[b], G. Saint-Girons[a] and R. Bachelet[a*]

[a] *Institut des Nanotechnologies de Lyon (INL) – CNRS UMR 5270, Ecole Centrale de Lyon, Bâtiment F7, 36 av. Guy de Collongue, 69134 Ecully Cedex, France. *E-mail: romain.bachelet@ec-lyon.fr*

[b] *Institut Lumière Matière (ILM) - CNRS UMR 5306, UCBL, Bât. Léon Brillouin, Campus LyonTech La Doua, 69622 Villeurbanne Cedex, France.*

[c] *Institut d'Electronique et des Systèmes (IES), CNRS, Universite Montpellier 2, 860 rue de Saint Priest, 34095 Montpellier, France.*

[d] *Institut de Ciencia de Materials de Barcelona (ICMAB -CSIC) Campus UAB, E-08193 Bellaterra, Catalunya, Spain.*

[*] E-mail: romain.bachelet@ec-lyon.fr

**Thermoelectric La-doped $SrTiO_3$ epitaxial layers with single-crystal quality: from nm to μm and mosaicity effects**

**Graphical Abstract**

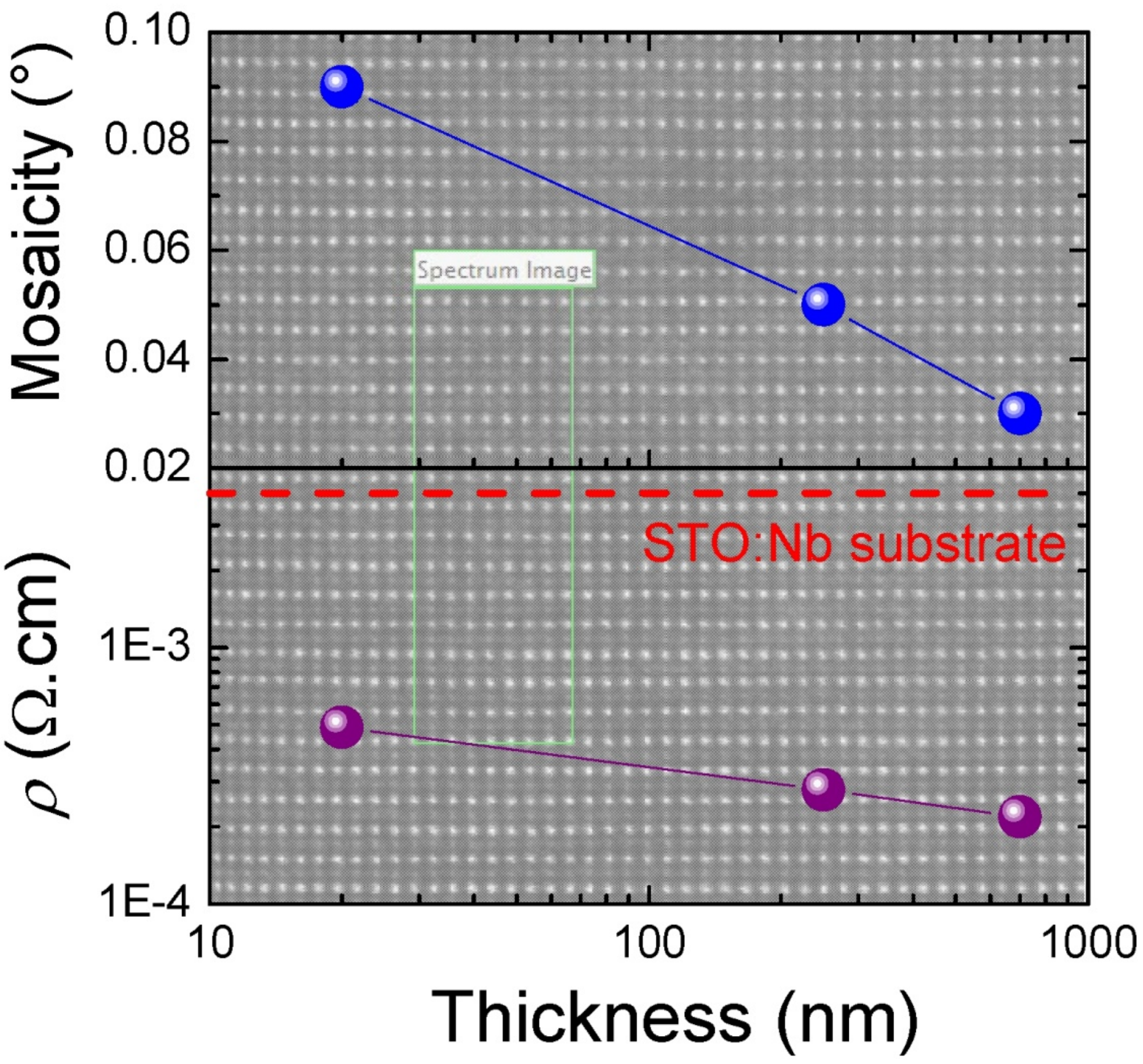

# Thermoelectric La-doped $SrTiO_3$ epitaxial layers with single-crystal quality: from nm to μm and mosaicity effects

**Abstract:** High-quality thermoelectric $La_xSr_{1-x}TiO_3$ (LSTO) layers (here with x = 0.2), with thicknesses ranging from 20 nm to 0.7 μm, have been epitaxially grown on $SrTiO_3$(001) substrates by enhanced solid-source oxide molecular-beam epitaxy. All films are atomically flat (with rms roughness < 0.2 nm), with low mosaicity (<0.1°), and present very low electrical resistivity (<5 x $10^{-4}$ Ω cm at room temperature), one order of magnitude lower than commercial Nb-doped $SrTiO_3$ single-crystalline substrate. The conservation of transport properties within this thickness range has been confirmed by thermoelectric measurements where Seebeck coefficients of around -60 μV/K have been found for all films, accordingly. Finally, a correlation is given between the mosaicity and the (thermo)electric properties. These functional LSTO films can be integrated on Si in opto-microelectronic devices as transparent conductor, thermoelectric elements or in non-volatile memory structures.

## 1. Introduction

Functional oxides present high chemical and thermal stability with large degree of compositional freedom, which offers great flexibility to tune their physical properties.[1,2] Therefore, functional oxides containing nontoxic, abundant, and cheap elements, are alternative and viable candidates for many devices in large-scale application fields such as non-volatile memories, sensors, actuators, energy harvesters. Especially, perovskite oxides, of general formula $ABO_3$, exhibit a wide range of properties such as ferroelectricity, piezoelectricity, pyroelectricity, ferromagnetism, thermoelectricity, metal-insulator transitions.[1,2] One of the most known $ABO_3$ archetype is $SrTiO_3$ (STO), a dielectric material at room temperature, which displays, for instance, tunable ferroelectric transition by isovalent $Ba^{2+}$ substitution on A-Site,[3] and tunable insulator-metallic (MI) transition by substituting $La^{3+}$ on A-site, $Nb^{5+}$ on B-site or by introducing oxygen vacancies.[4] In this light, specific $La^{3+}$ doping level in $SrTiO_3$, $La_xSr_{1-x}TiO_3$ (LSTO), can present high electrical conductivity (up to ~$10^4$ S/cm at room temperature), large Seebeck coefficient (up to ~1 mV/K at room temperature), large thermoelectric power factors (up to ~40 $\mu$W $cm^{-1}$ $K^{-2}$ at room temperature, larger than the thermoelectric $Bi_2Te_3$-family materials) [4,5] as well as large electronic mobility (up to ~$10^4$ $cm^2$ $V^{-1}$ $s^{-1}$ at low temperature) [6] and optical transparency in the visible range while maintaining metallic conductivity.[7] LSTO could therefore be used in important applications as transparent conductor,[7] thermoelectric element,[5] or non-volatile memory involving MI transitions or field effects.[8-10] However, its physical properties strongly depend on several parameters during the elaboration process that impacts on its atomic and chemical structure. Since applications require the use of high-quality films, epitaxial growth of single-crystalline films of controlled properties over a wide thickness range is a prerequisite. However, the results published so far on epitaxial LSTO films are restricted to a few nanometers thick films with incomplete experimental or structural details.

Nanometric epitaxial LSTO films have been grown mainly using pulsed laser deposition (PLD) [7,11,12] and a particular hybrid molecular beam epitaxy (MBE) technique involving an organo-metallic precursor as Ti source.[5,6] Since oxygen vacancies act as an additional *n*-type dopant,[7] the transport properties of the La-doped STO films can be altered by means of different oxygen partial pressures $P(O_2)$ during deposition.[5,7,11] The LSTO grown by PLD can have metallic behavior with resistivities ($\rho$) of the order of $10^{-3}$ $\Omega$ cm. For instance $\rho = 4.6$ m$\Omega$ cm has been reported with 15% La doping in films grown at $P(O_2) = 10^{-3}$ Torr.[7] Decreasing $P(O_2)$ below $10^{-5}$ Torr results in a resistivity drop by a factor of ~2 due to the introduction of oxygen vacancies, in agreement with the increased carrier density.[11,12] Electrical resistivity of nanometric LSTO films grown by MBE is of about one order of magnitude lower than that of films grown by PLD, taking into account the same La doping level and the same substrate.[7,13] For instance, with 15% La doping, a resistivity down to a few $10^{-4}$ Ω cm can be reached by MBE at room temperature even after oxygen annealing of the samples and consequent suppression of oxygen vacancies.[13] This difference can be due to the non-equilibrium energetic character of PLD process, the higher residual background pressures of the PLD deposition chambers, and the non-obvious stoichiometric transfer from the target to the film due to the dependence on laser fluence and difference in volatility or sticking coefficients between elements.[14] Non stoichiometric transfer is known to occur at least with Pb-, Bi-, K-, Ru-based oxides,[15,16] and even with La-based oxides being one of the possible reason for observing *n*-type conductivity at the $LaAlO_3/SrTiO_3$ interface.[17] Therefore, MBE appears as a reference equilibrium high-purity elaboration technique for the growth of state-of-the-art conducting oxide films with controlled composition. However, most of the MBE-grown LSTO films are elaborated with nanomoetric thickness (<150 nm), by a particular hybrid MBE method, or for which elaboration or structural details are elusive.[5-6,13] A possible explanation is that the main challenge in solid-source oxide MBE is the stability and reproducibility of the fluxes, controlling the stoichiometry. Here, epitaxial LSTO single-layers are grown by enhanced solid-source oxide MBE up to ~µm in thickness, maintaining excellent structural quality (low mosaicity down to 0.03°) and low electrical resistivity ($< 5 \times 10^{-4}$ Ω cm at room temperature).

Additionally, growth conditions with the correlated structural and functional properties of the layers are described in full details.

## 2. Experimental details

Epitaxial $La_{0.2}Sr_{0.8}TiO_3$ (LSTO) films were grown on (001)-oriented $SrTiO_3$ (STO) substrates by solid source oxide MBE using effusion cells, metallic elements and molecular oxygen. This elaboration technique allows the monolithic integration of perovskite films on Si *via* a $SrTiO_3$ buffer-layer that assists the fabrication of oxide-based devices for the microelectronic industry.[18-19] The crucibles used in the effusion cells are made of tungsten for high-temperature refractory-Ti evaporation (promoting its challenging flux stability), pyrolytic boron nitride (PBN) for Sr and tantalum for La. Net fluxes were measured by a Bayard-Alpert ionization gauge with Thoria-coated Ir filament located at the sample position in the center of the chamber, subtracting the residual background pressures at different temperatures (see Fig. S1 in supplementary material). A low residual base pressure of about a few $10^{-10}$ Torr is achieved in the MBE chamber by using combined turbo-molecular and liquid nitrogen cryogenic pumping. Before growth, the substrates were annealed for 10 minutes under a di-oxygen partial pressure of $10^{-7}$ Torr, in the MBE chamber, at a temperature of 600 °C. The epitaxial growth has been performed in co-deposition process at a constant substrate temperature of 450 °C, using a di-oxygen partial pressure of $10^{-7}$ Torr (avoiding oxidation of the metallic sources and thus flux deviations) and a growth rate of ~1.5 ML/min. Growth rate and film thickness estimation have been done using x-ray reflectometry (XRR) measurements and cross-checked by transmission electron microscopy. Film thickness ranges here from 20 nm to 0.7 µm. I*n-situ* reflection high-energy electron diffraction (RHEED) was employed to monitor the structural quality of the film during the growth. The LSTO films were further annealed ex-situ during a few hours at 450 ˚C under air in a tubular furnace in order to eliminate the oxygen vacancies in both films and substrates and to ensure reliability of the transport properties measurements.[20] The substrates were then checked to be well insulating. The film morphology was analyzed using atomic force microscopy (AFM) in tapping mode. Their crystalline quality was investigated using X-ray diffraction (XRD) (SmartLab diffractometer from Rigaku equipped with a high brilliance 9 kW X-ray source rotating anode, two-bounce Ge(220) monochromator, and two crossed $R_x$-$R_y$ cradles enabling precise alignment). The stoichiometry of the LSTO films was determined by X-ray photoelectron spectroscopy (XPS). One representative exemple of the core level spectra is illustrated in Fig. S2 in supplementary material. One film's microstructure was examined with a NION UltraSTEM scanning transmission electron microscope (STEM) operated at 100 kV and equipped with a NION aberration corrector and a Gatan Enifina dedicated spectrometer for electron energy loss spectrometer (EELS) experiments. Electrical resistivity measurements were performed in Van der Pauw geometry from room to low (10 K) temperature by using a 4-Watt ARS displex refrigeration system. For this purpose, ultrasonic wire bonding (West-Bond machine) was used to connect aluminum wires to Cu bonding pads on the corners of the squared samples. Seebeck coefficient measurements using the differential method (see Fig. S3 in supplementary material) were performed from room temperature to ~100 K using a similar device as the one described in Ref. [21].

## 3. Results and discussion

LSTO epitaxial films of different thicknesses, from 20 nm up to 0.7 μm, were grown on (001)-oriented STO substrates. The RHEED patterns of the as-grown LSTO films are shown in Figure 1(left). For all studied thicknesses, RHEED patterns present high diffraction contrast with entire-order reflection streaks along the <110> and <100> directions after growth, indicating that the films are epitaxial with stoichiometric and atomically-flat surface.[22] This is confirmed by the AFM topographic images (Fig. 1, right), where root-mean-square (rms) roughness less than 0.2 nm and peak-to-peak height amplitude less than one unit-cell (~0.4 nm) are measured for all film thicknesses up to 0.7 μm.

The structural properties of the LSTO films have been measured by XRD (Fig. 2). Apart the intense Bragg diffraction peaks corresponding to the {00l} reflections of the LSTO film and STO substrate, no other reflections were observed along the [00l] out-of-plane direction, showing that the films are well single-oriented. The LSTO (002) reflection of the 20 nm thick film shows Pendellösung fringes, indicating good crystalline quality with atomically-flat interfaces (Fig. 2a). As expected, the diffracted intensity increases with the thickness of the film and the peaks width decreases. The measured out-of-plane lattice parameters ($c$) were found to be near 3.935 Å for all films. The variation of the lattice parameter is within 3% stoichiometry deviation.[14] This value depends on the partial oxygen pressure during deposition, epitaxial strain and doping concentration, and it is in agreement with reported values by other groups.[11-13] Importantly, a full width at half maximum (FWHM) below 0.1˚ of the (002) reflection rocking curve of the 20 nm thick film (inset of Fig. 2a) demonstrates the low mosaicity of the film, better than other groups.[7,13] Furthermore, the mosaicity value decreases when increasing the film thickness, reaching a lower value of 0.03° for a LSTO film of 0.7 μm thick (inset of Fig. 2c) which is close to that of a STO single-crystalline substrate.

To illustrate the high structural quality of the LSTO films, a STEM characterization has been done on a 250 nm thick film (Fig. 3). Figure 3(a) shows a Z-contrast image which reveals a defect-free and continuous over a long lateral length LSTO layer. A higher magnification Z-contrast image of the lattice shows a perfect LSTO crystallinity, with no structural distortions, neither in the La/Sr nor in the Ti sublattices (Fig. 3b). Atomically resolved EELS spectrum images were also acquired (Fig. 3(c-f)) to study any possible chemical ordering. The elemental maps show that the $ABO_3$ lattice is preserved, with the Sr substitution by La in the A-site of the perovskite structure. Although La may seem to be not homogeneously distributed at the atomic scale, different EELS spectrum images acquired from different sample regions have shown that the La doesn't form clusters and that is globally well-distributed within the layer.

Figure 4(a-c) presents the electrical resistivity of the LSTO films as a function of temperature for all the thickness range. A 0.7 wt% Nb-doped STO single-crystalline substrate has been measured for comparison (Fig. 4d). All the LSTO films behave as metallic conductors throughout the entire thickness range, with resistivities below 5 x $10^{-4}$ Ω cm at room temperature, of about one order of magnitude lower than commercial 0.7 wt% Nb-doped STO single-crystalline substrates (4 x $10^{-3}$ Ω cm, in agreement with the providers data sheets). The temperature dependence exhibits the characteristic $T^2$-dependence of a Fermi liquid as previously observed experimentally in doped-STO over a wide temperature and doping ranges.[5,7,23,24] The resistivity values of our LSTO films with 20% La doping (3 x $10^{21}$ $cm^{-3}$ carrier concentration) are consistent with the reports of other groups, depending on the doping concentration and strain state.[5,13] The slight decrease of resistivity with thickness here can be explained by the decrease of the mosaicity with thickness (see below, Figure 6).

The Seebeck coefficients (S), $S = \Delta V/\Delta T$ where $\Delta V$ is the thermo-emf induced by the thermal gradient $\Delta T$ (see Fig. S3 in supplementary material), were measured as function of temperature, from room temperature down to 100-150 K (Fig. 5). The measured Seebeck coefficients at room temperature of all the films are around -60 μV/K, in perfect agreement with the doping level, electrical resistivity and previous reports.[5] Consistently, the slight measured S difference is inversely proportional to the slight difference in resistivity values. It is worth to note that the thermoelectric properties of doped STO was recently very well reproduced within four orders of magnitude for electrical conductivity and Seebeck coefficient in the context of a theoretical tight-binding approach based on a fully realistic band Hamiltonian, where only the electron-electron scattering process is considered.[25] Finally, a comparison of the resistivity and Seebeck coefficients with mosaicity is shown as a function of the thickness in Figure 6. The decrease of the resistivity and Seebeck coefficient is correlated with the decrease of the mosaicity. The mosaicity appears as another possible parameter to tune the transport properties of La-doped $SrTiO_3$ films.

## 4. Conclusions

Highly-conductive thermoelectric $(La_{0.2}Sr_{0.8})TiO_3$ epitaxial films with excellent structural quality have been grown by solid-source oxide MBE up to ~μm thickness range. All the films have atomically-flat surface and low mosaicity (< 0.1°) and present electrical resistivity ($< 5 \times 10^{-4}$ Ω cm), one order of magnitude lower than commercial Nb-doped STO single-crystals. Seebeck coefficient measurements have confirmed the electrical properties. STEM images have shown the high structural quality and homogeneity of the film. The slight variations of the transport properties are attributed here by the slight variations of the mosaicity. These results underline the potential of La-doped STO films grown by standard MBE technique for applications as transparent conductor or thermoelectric element where specific doping concentration or μm thick integrated layers are required.

**Acknowledgements**

The european commission is acknowledged for the funding of the project TIPS (H2020-ICT-02-2014-1-644453). J.G. acknowledges the Ramon y Cajal program (RYC-2012-11709). Part of this research (STEM characterizations) was conducted at the Center for Nanophase Materials Sciences (Oak Ridge National Laboratory), which is a DOE Office of Science User Facility. The joint laboratory RIBER-INL and J.-B. Goure are finally acknowledged for the MBE technical support.

## Figure captions

**Figure 1**. (left) RHEED patterns along the <100> and <110> directions after growth, and (right) AFM topographic images and corresponding topographic profiles of (a) 20 nm thick, (b) 250 nm thick and (c) 0.7 µm thick LSTO layers.

**Figure 2**. XRD θ-2θ symmetrical scans, and ω-scans around the (002) LSTO reflection in inset, of (a) 20 nm thick, (b) 250 nm thick and (c) 0.7 µm thick LSTO layers.

**Figure 3**. (a) Low magnification and (b) high magnification Z-contrast images of a 250 nm thick LSTO layer viewed along the STO[110] zone axis. Scale bars represent 200 nm in (a) and 4 nm in (b). (c) Annular-dark field (ADF) image of the region marked in (b). (d-f) EELS Elemental maps corresponding to Ti $L_{2,3}$, La $M_{4,5}$ edges, and Sr $L_{2,3}$, respectively.

**Figure 4**. Temperature dependence of the electrical resistivity of (a) 20 nm thick, (b) 250 nm thick and (c) 0.7 µm thick LSTO layers, and (d) single-crystalline 0.7 wt% Nb-doped STO(001) substrate.

**Figure 5**. Temperature dependence of the Seebeck coefficient of (a) 20 nm thick, (b) 250 nm thick and (c) 0.7 µm thick LSTO layers.

**Figure 6**. Overview of the structural and transport properties at 300 K against thickness of the LSTO layers: (a) mosaicity, (b) electrical resistivity, and (c) Seebeck coefficient.

**Figure 1**

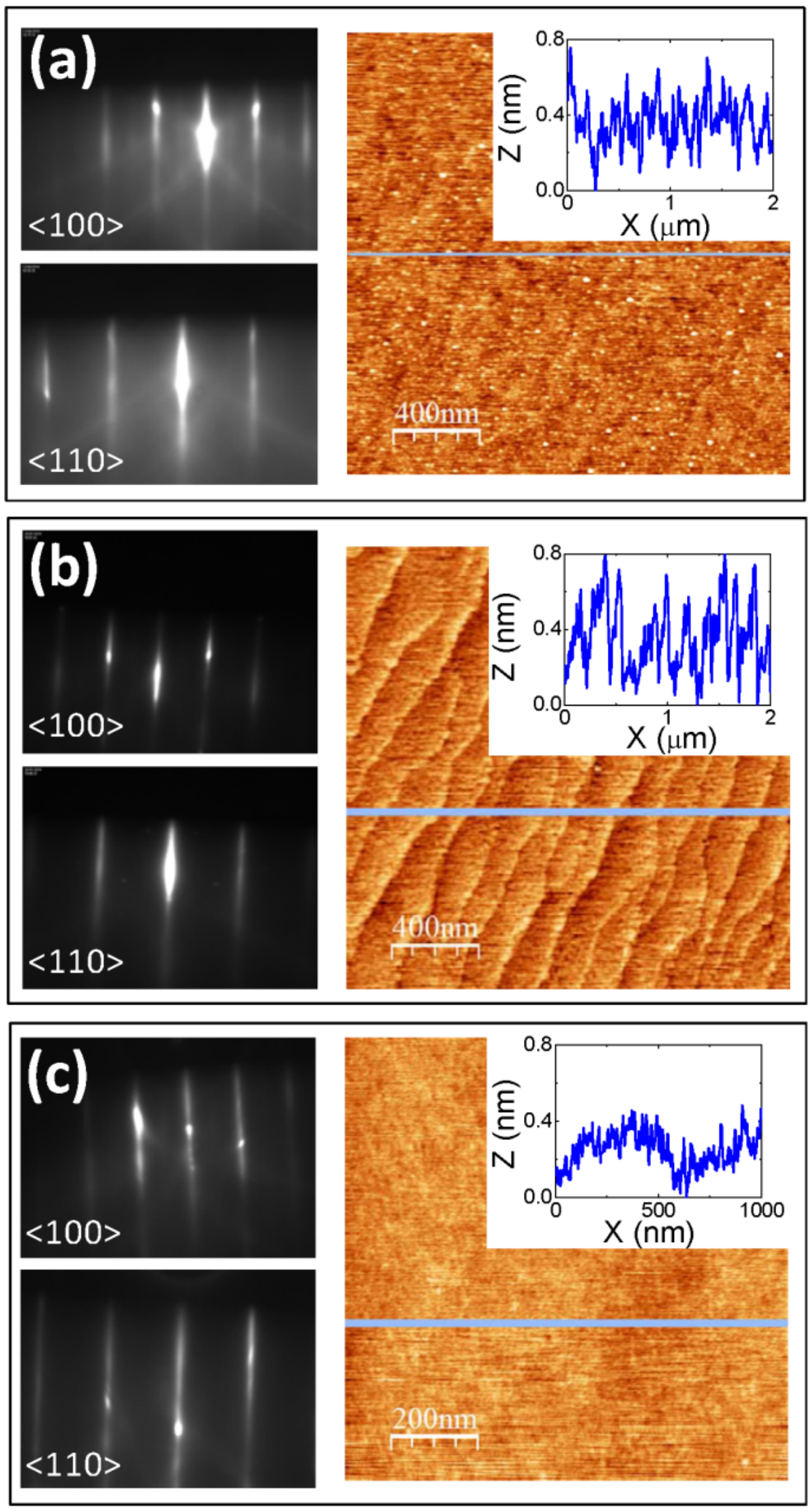

**Figure 2**

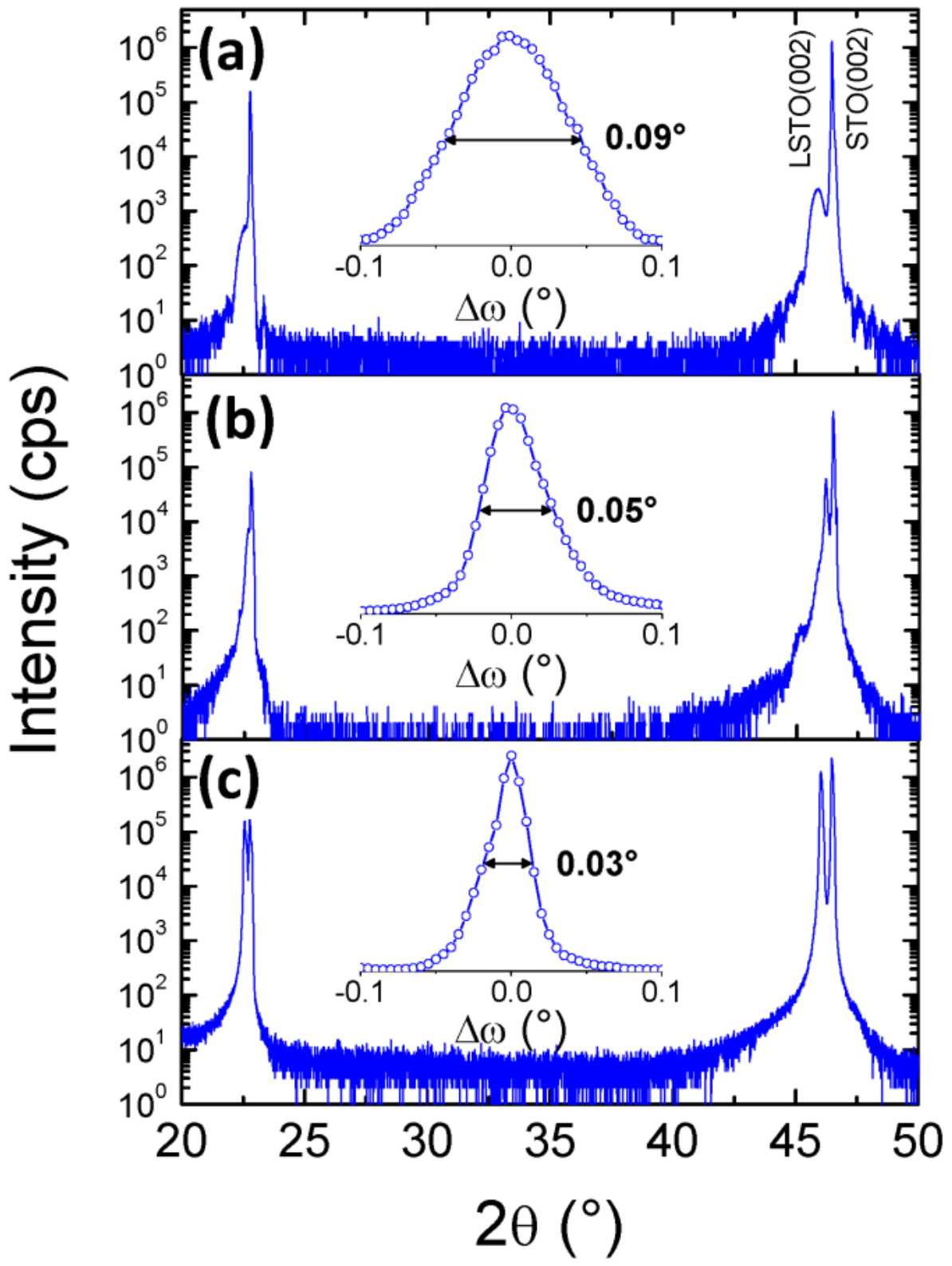

**Figure 3**

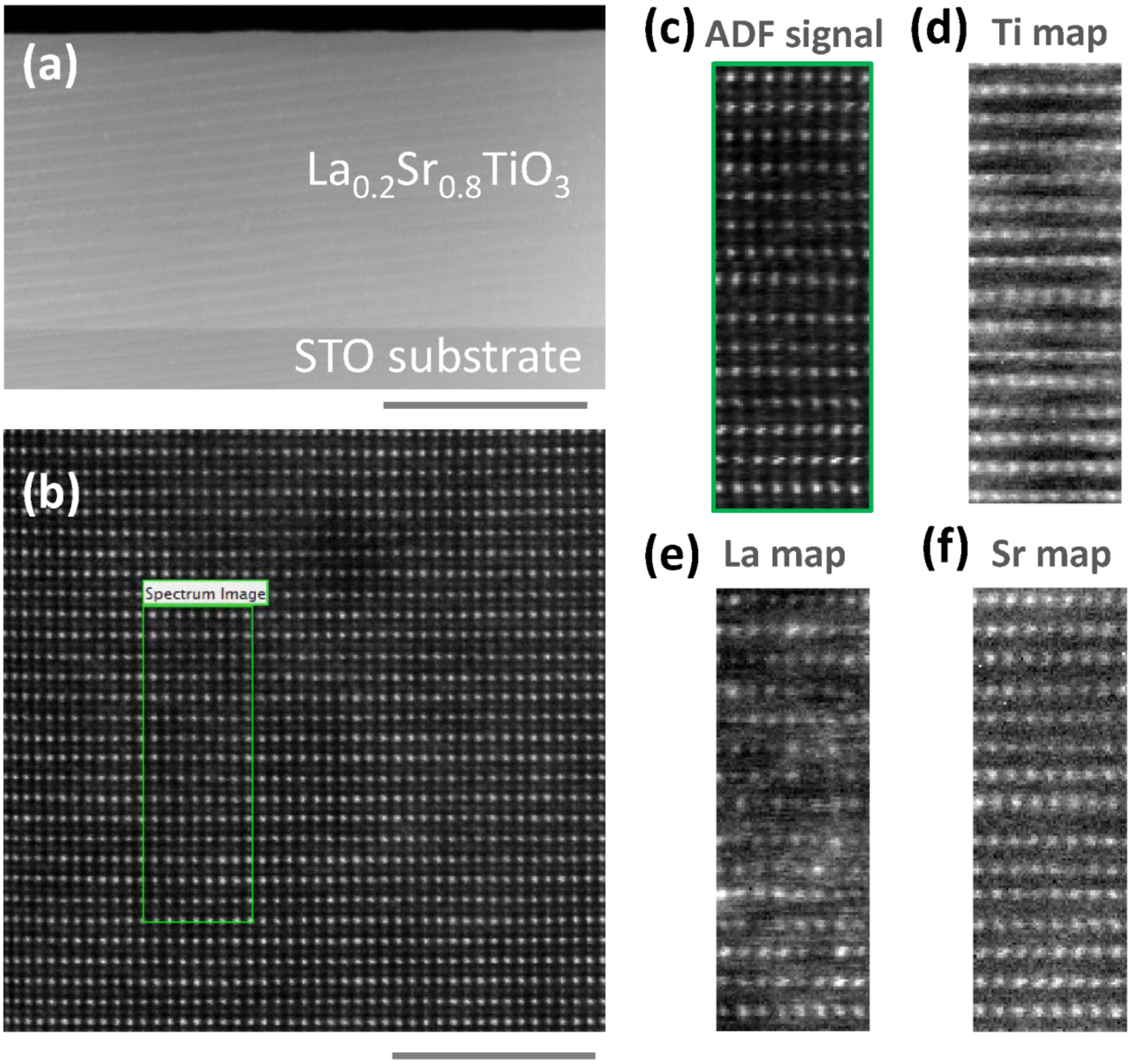

**Figure 4**

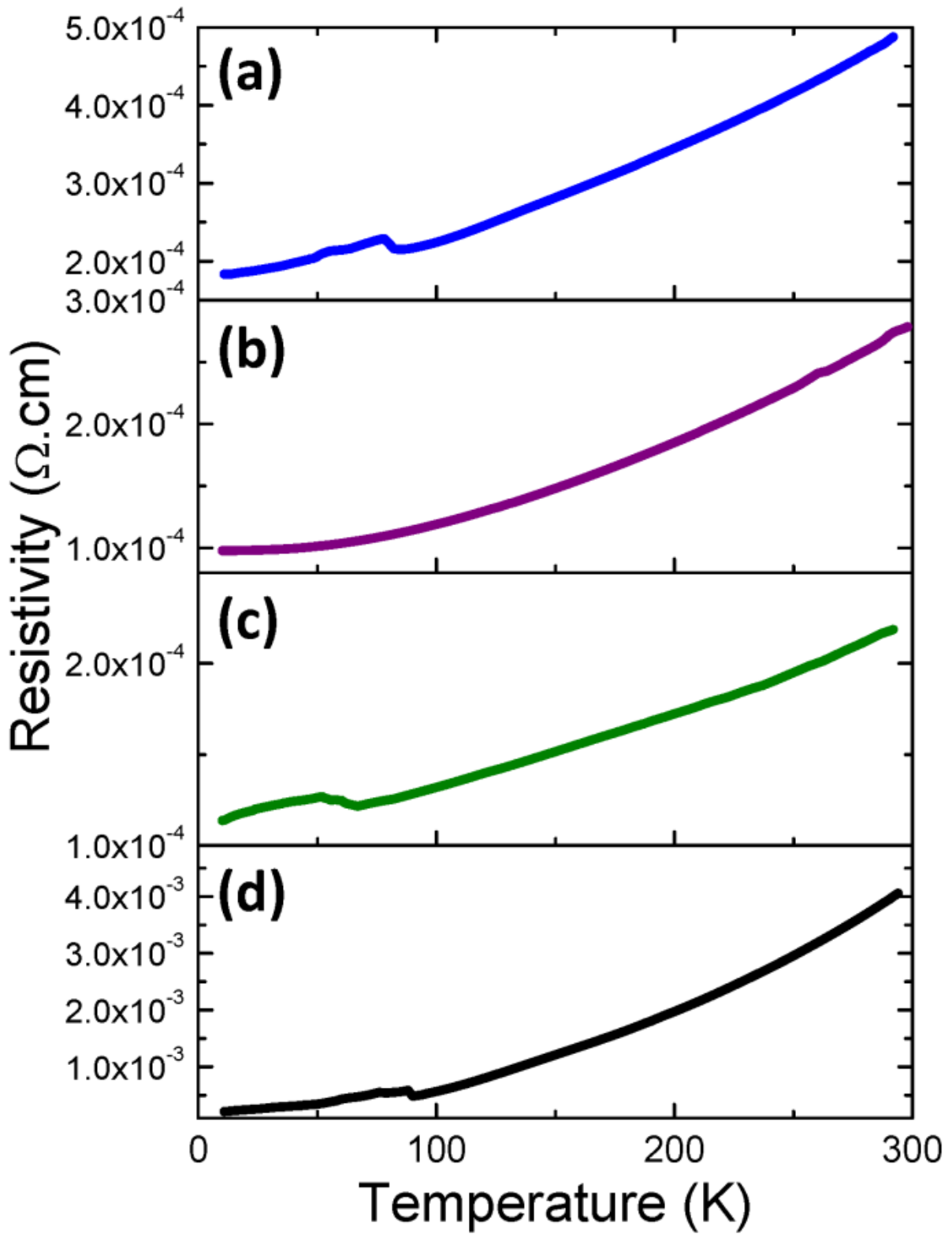

**Figure 5**

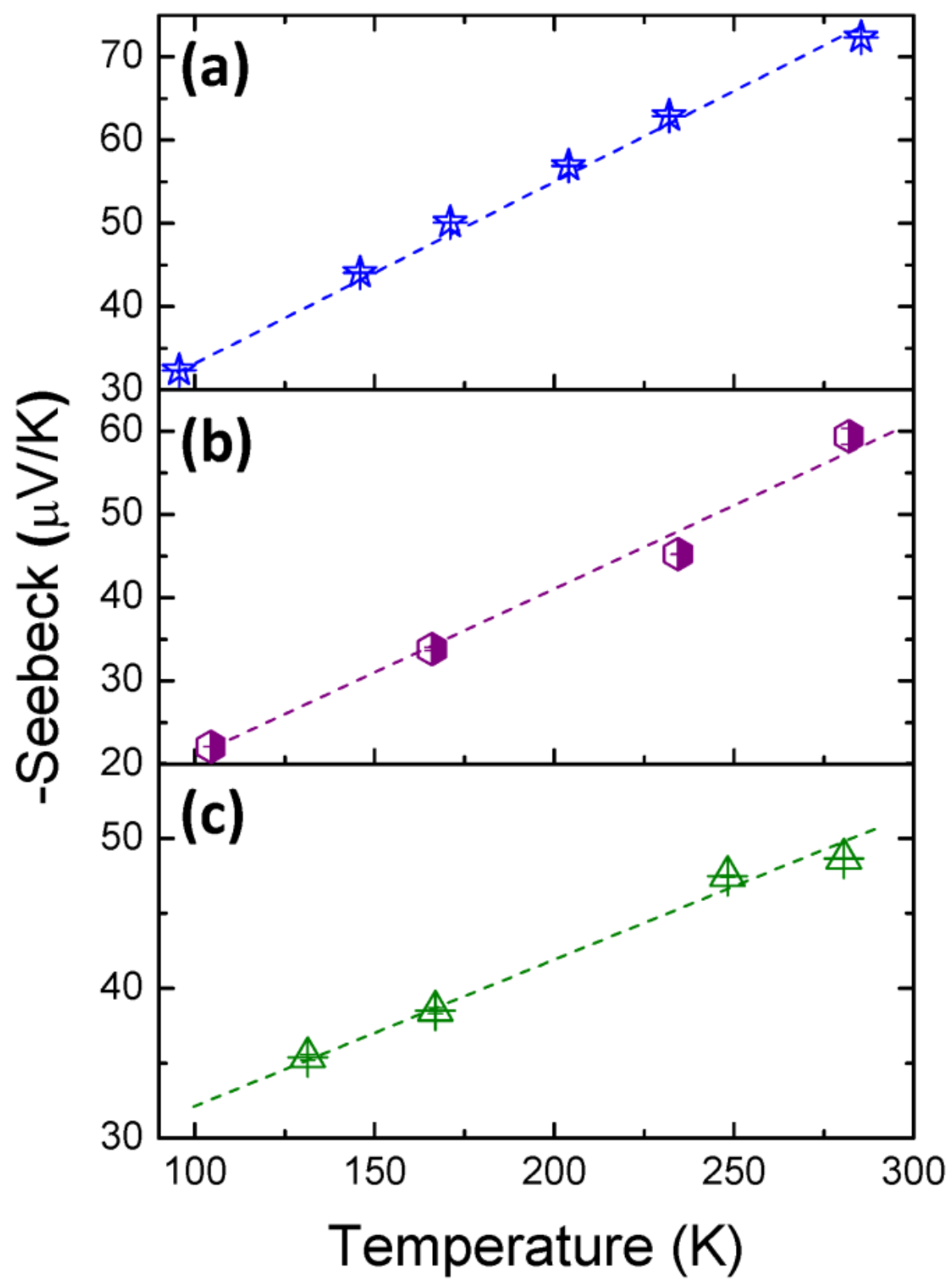

**Figure 6**

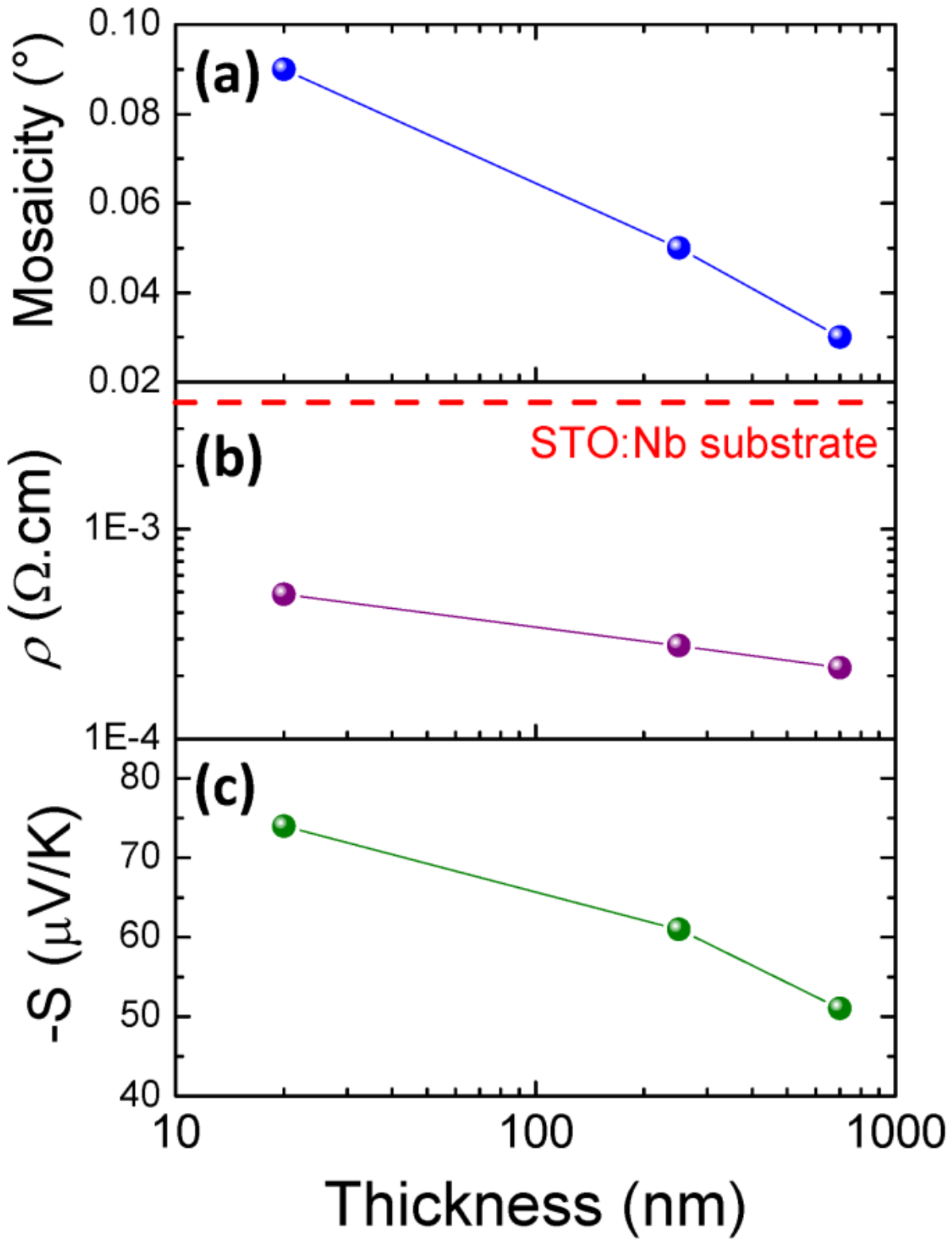

# Supplementary Material

## Content of the Supplementary Material:

- MBE net fluxes measured by Bayard-Alpert gauge for each metallic element (Fig. S1)
- XPS spectra of the La4d, Sr3d, Ti2p and O1s core levels (Fig. S2)
- Seebeck coefficients measurements around room temperature (Fig. S3)

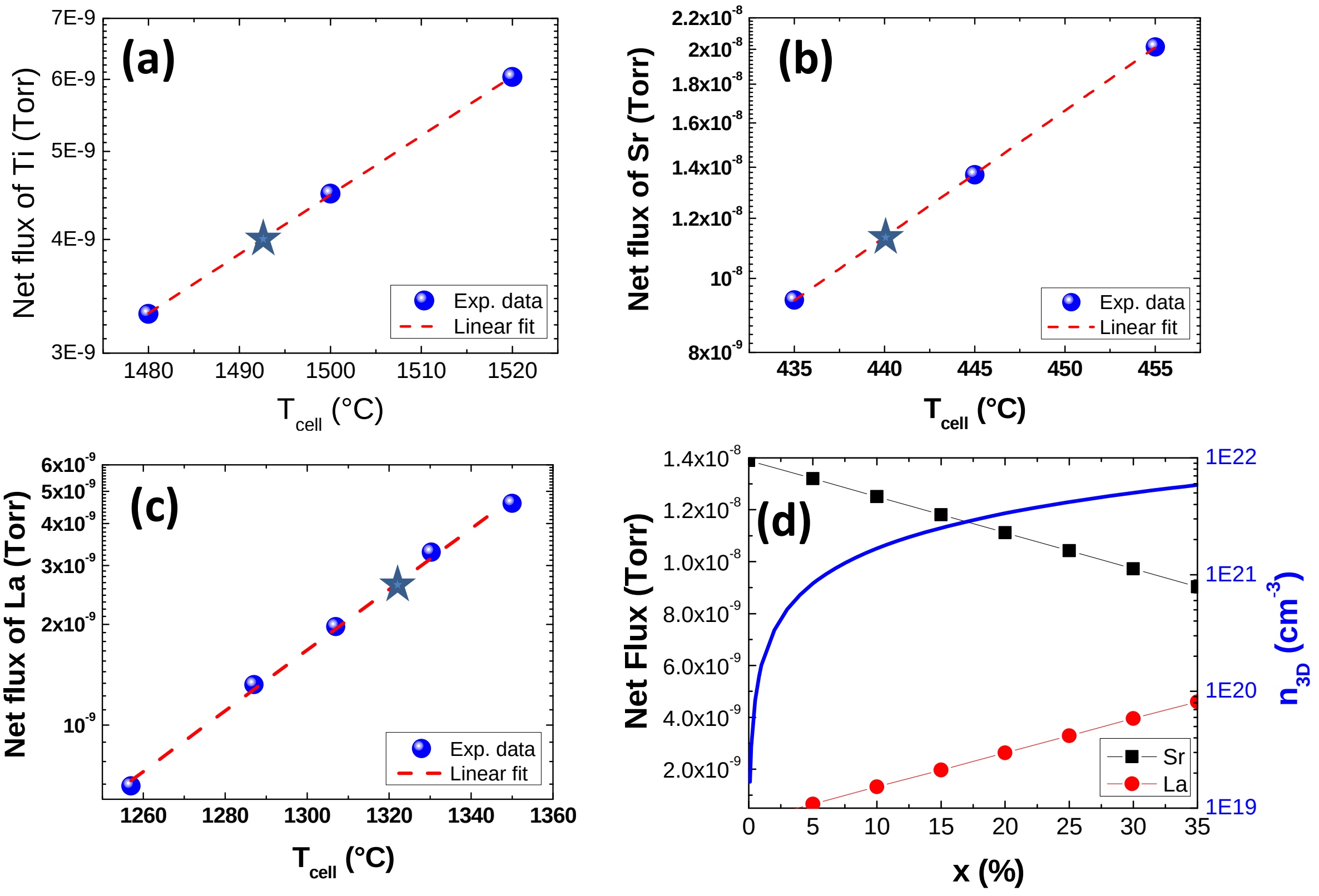

**Figure S1:** MBE net fluxes as a function of temperature of (a) Ti, (b) Sr, and (c) La effusion cells. (d) MBE net fluxes for La and Sr, as well as the expected carrier concentration ($n_{3D}$), as a function of the La doping concentration (x) in $La_xSr_{1-x}TiO_3$ (LSTO). The star points in (a-c) represent the net fluxes and corresponding effusion cell temperatures used in this paper for x=0.2.

Figure S1 shows the net flux for each metallic element (measured using the Bayard-Alpert gauge under UHV and after subtraction of the background pressure) at different effusion cells temperatures in the target composition range ($La_{0.2}Sr_{0.8}TiO_3$). The logarithm of the net flux increases linearly with temperature, as expected [S1-S2]. Net flux as small as a few $10^{-10}$ Torr (in the order of magnitude of the residual base pressure) can be finely measured with the procedure detailed above. The net flux of Ti is set to 4 x $10^{-9}$ Torr, corresponding to a cell temperature of around 1500 °C and to a growth rate of 1.5 ML/min (Fig. S1a). The corresponding 20% La doping and consequent 80% Sr, to maintain a cationic stoichiometry ratio A/B = 1, is obtained at a La net flux of 2 x $10^{-9}$ Torr for a temperature around 1310°C, and a Sr net flux of 1.1 x $10^{-8}$ Torr for a temperature of ~440°C, respectively (Fig. S1(b-d)). The overall chemical composition of LSTO films were estimated using XPS measurements where relative elemental concentrations were obtained, crosschecked with surface reconstructions by *in-situ* RHEED [S3]. Based on these results, the working temperatures of each effusion cell were adjusted accordingly to obtain the desired composition.

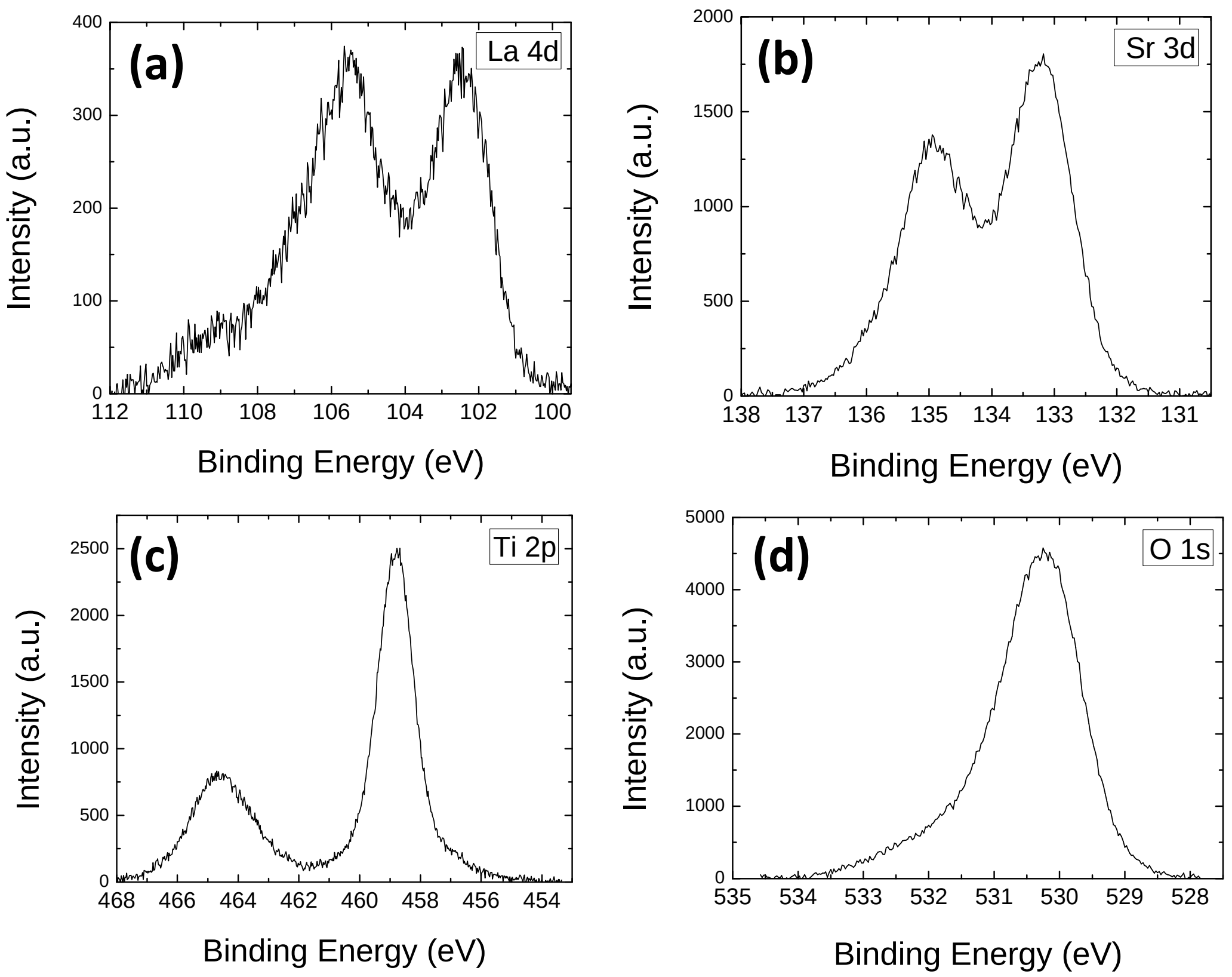

**Figure S2:** Representative XPS spectra of (a) La 4d, (b) Sr 3d, and (c) Ti 2p, and (d) O 1s core levels, taken here on the 0.7 µm thick LSTO film after air annealing at 450°C.

Figure S2(a-d) displays typical XPS spectra of La 4d (a), Sr 3d (b), Ti 2p (c) and O 1s (d) core levels, taken here on the thickest LSTO film after air annealing. The peak positions correspond to well loxidized LSTO. The low value of full-width at half-maximum (FWHM) of the O 1s core level peak (1.3 eV) attests the good crystalline quality of the LSTO layer. Besides, the carrier concentration was estimated based on the XPS results and confirmed by Hall effect measurements. It was found that the La donors are ionized and each La donates one electron to the conduction band of $SrTiO_3$ as expected in this doping range [S4].

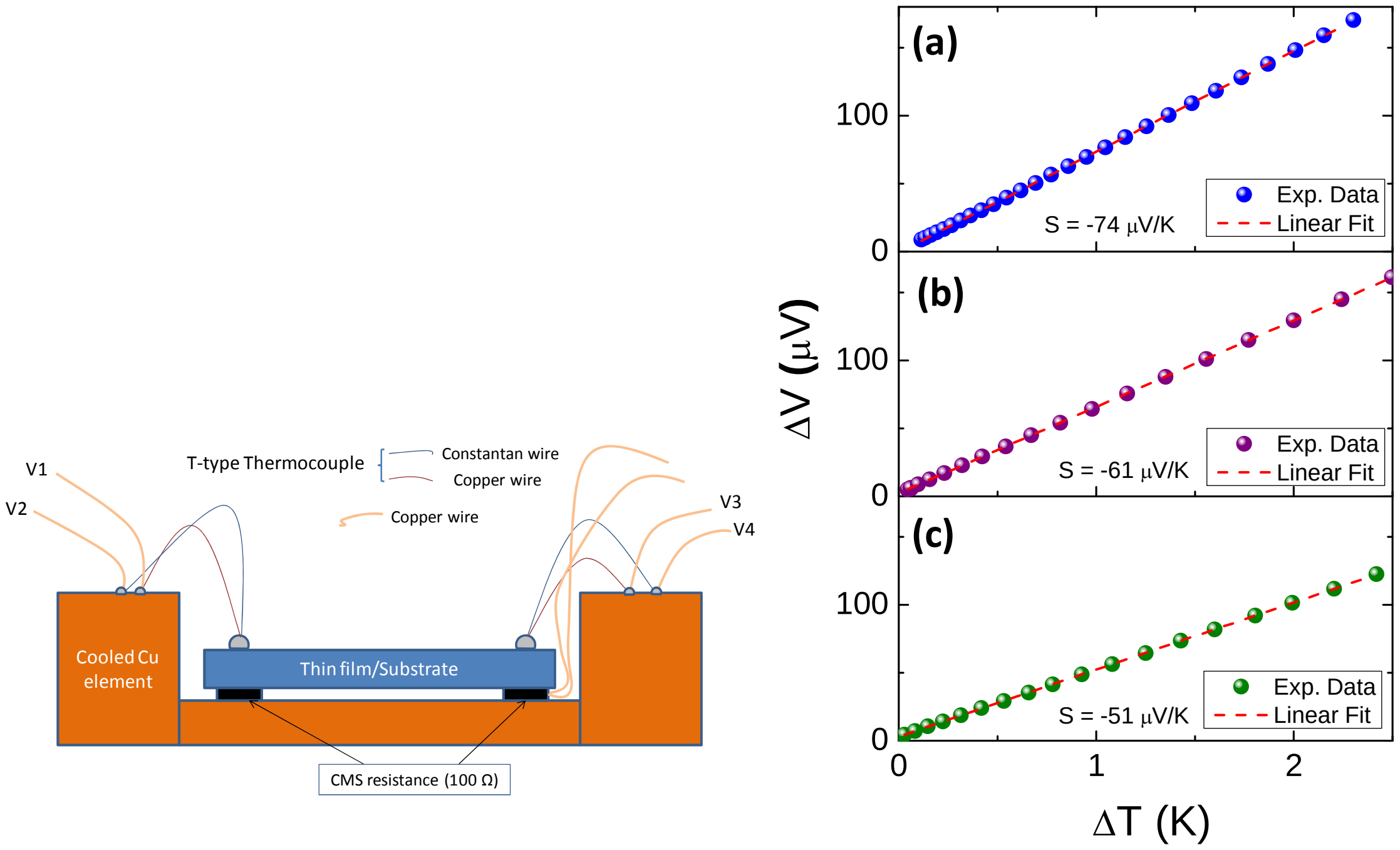

**Figure S3:** (left) Sketch of the set-up used for Seebeck coefficient (S = ΔV/ΔT) measurements in our films, and (right) S measurements around room temperature of (a) 20 nm, (b) 250 nm, and (c) 0.7 μm thick LSTO films.